\documentclass[pre,amsmath]{revtex4}
\usepackage[usenames]{color}
\usepackage{graphicx}
\begin{document}

\title{Force-induced breakdown of flexible polymerized membrane}

\author{J. Paturej$^{1,2}$, H. Popova$^3$ A. Milchev$^{1,3}$, and T.A. Vilgis$^1$}

\affiliation{
$^1$ Max Planck Institute for Polymer Research, 10 Ackermannweg, 55128 Mainz,
Germany\\
$^2$ Institute of Physics, University of Szczecin, Wielkopolska 15, 70451
Szczecin, Poland\\
$^3$ Institute of Physical Chemistry, Bulgarian Academy of Sciences, 1113 Sofia,
Bulgaria}

\begin{abstract}
We consider the fracture of a free-standing two-dimensional (2D) elastic-brittle
network to be used as protective coating subject to constant tensile stress
applied on its
rim. Using a Molecular Dynamics simulation with Langevin thermostat, we
investigate the scission and recombination of bonds, and the formation of cracks
in the 2D graphene-like hexagonal sheet for different pulling force $f$ and
temperature $T$. We find that bond rupture occurs almost always at the sheet
periphery and the First Mean Breakage Time $\langle \tau \rangle$ of bonds
decays with membrane size as $\langle \tau \rangle \propto N^{-\beta}$ where
$\beta \approx 0.50\pm 0.03$ and $N$ denotes the number of atoms in the
membrane. The
probability distribution of bond scission times $t$ is given by a Poisson
function $W(t) \propto t^{1/3} \exp (-t / \langle \tau \rangle)$. The mean
failure time $\langle \tau_r \rangle$ that takes to rip-off the sheet declines
with growing size $N$ as a power law $\langle \tau_r \rangle \propto
N^{-\phi(f)}$. We also find $\langle \tau_r \rangle \propto \exp(\Delta
U_0/k_BT)$ where the nucleation barrier for crack formation $\Delta U_0 \propto
f^{-2}$, in agreement with Griffith's theory.  $\langle\tau_r\rangle$ displays
an Arrhenian dependence
of $\langle \tau_r \rangle$ on temperature $T$. Our results indicate a rapid
increase in crack spreading velocity with growing external tension $f$.
\end{abstract}

\maketitle

\section{Introduction}

Fracture in engineering materials is a long-standing topic of research due to
problems  that arise with technological applications and the ensuing economical
implications. Thus, for decades a lot of attention has been focused on
understanding the macroscopic and microscopic factors which trigger failure.
Recently, the interest and the need for better understanding of the interplay
between elastic and fracture properties of brittle materials  has been revived
due to the rapidly developing design of advanced structural materials.

Promising aspects for applications include reversible polymer networks
\cite{Sijbesma1,Sijbesma2}, and also graphene, that shows unusual
thermomechanical properties \cite{Peeters,Aluru}.
%Graphene - defined as the monolayer of honeycomb lattice packed with carbon
%atoms - has gained significant attention recently not only due to its unusual
%electronic properties but also in view of its exceptional mechanical
%performance
%\cite{Hone} which makes it a prominent candidate for potential applications
%as micro- and nano-electro-mechanical device. 
Among other things, graphene, which is a honey-comb lattice packed with $C$
atoms
can be used as anti-corrosion gas barrier protective coating \cite{Ruoff}, in
chemical and bio-sensors \cite{Pumera},  or as efficient membrane 
for gas separation \cite{Jiang}. In all possible applications the temperature
and
stress-dependent fracture strength of this 2D-network is of crucial importance.
Graphene has been investigated recently by Barnard and Snook \cite{Snook} using
{\em ab initio} quantum mechanical techniques whereby it was noted that that the
problems ``has been overlooked by most computational and theoretical studies''.

An important example of biological microstructure is {\em spectrin}, the red
blood cell membrane skeleton, which reinforces the cytoplasmic face of the
membrane. In erytrhrocytes, the membrane skeleton enables it to undergo large
extensional deformations while maintaining the structural integrity of the
membrane. A number of studies, based on continuum- \cite{Skalak}, percolation-
\cite{Srolovitz,Saxton,Seifert}, or molecular level \cite{Monette,Dao}
considerations of the mechanical breakdown of this network, modeled as a
triangular lattice of spectrin tetramers, have been reported so far. Many of
these studies can be viewed in a broader context as part of the problem of
thermal decomposition of gels \cite{Argyropoulos}, epoxy resins
\cite{Rico,Zhang} and other 3D networks both experimentally
\cite{Argyropoulos,Rico,Zhang}, and by means of simulations \cite{Kober} in the
case of Poly-dimethylsiloxane (PDMS).

The afore-mentioned examples illustrate well the need for deeper understanding
of the processes of failure in brittle materials. Besides analytical and
laboratory investigations, computer simulations \cite{Holland,Tang,Cleri} have
provided meanwhile a lot of insight in aspects that are difficult for direct
observations or theoretical treatment - for a review of previous works see Alava
et al. \cite{Alava}. Most of these studies focus on the propagation of
(pre-existing) cracks, relating observations to the well known Griffith's model
\cite{Griffith} of crack formation. A number of important aspects of material
failure have found thereby little attention. Thus only a few simulations examine
the rate of crack nucleation which involves long time scales necessary for
thermal activation - see, however, \cite{Santucci,Wang,Benshaul,Gagnon}. Effects
of system size on the characteristic time for bond rupture have not been
examined except in a recent MD study by Dias et al. \cite{Grant}. Also
recombination of broken bonds has not been considered. These and other
insufficiently explored properties related to fracture have motivated our
present investigation of a free-standing 2D honeycomb brittle membrane by means
of Molecular Dynamics simulation. In view of the possible applications
as anti-corrosion and gas barrier coating, we consider a  radially-spanned
sheet of regular hexagonal flake shape so as to minimize effects of corners and
unequal edge lengths that are typical for ribbon-like sheets. Tensile constant
force is applied on the rim of the  flake, perpendicular to each edge. By
varying system size, tensile force and temperature, we collect a number of
results which characterize the initiation and the course of fragmentation in
stretched 2D honeycomb networks.

The paper is organized as follows: after a brief introduction, we sketch our
model in Sec. \ref{sec_model} where we consider interactions between atoms in
the brittle honeycomb membrane, define the threshold for bond scission, and also
introduce some basic quantities that are measured in the course of the
simulation. In Sec. \ref{sec_MD} we present our simulation results, presenting
briefly the results on recombination of broken bonds - \ref{subsec_rec}, the
distribution of bond scission rates over the membrane surface, the dependence of
the Mean First Breakage Time (MFBT) before a bond scission takes place and of
the mean failure time until the 2D sheet breaks apart on applied tensile force,
and examine how these times depend on membrane size and temperature -
\ref{subsec_MFBT}. The formation of cracks at different cases of applied stress
as well as their propagation in a 2D honeycomb brittle sheet are briefly
considered in subsection \ref{subsec_cracks}. We end this report by a brief
summary of results in Section \ref{sec_summary}.

\section{Model and Simulation Procedure}\label{sec_model}

\subsection{The model}

\begin{figure}[ht]
\begin{center}
\includegraphics[scale=0.69]{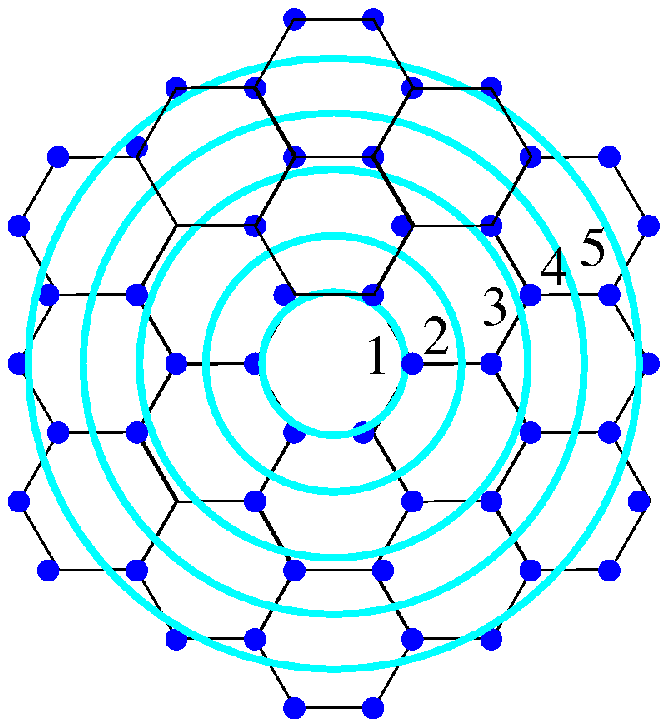}
\hspace{1.5cm}
\includegraphics[scale=0.55]{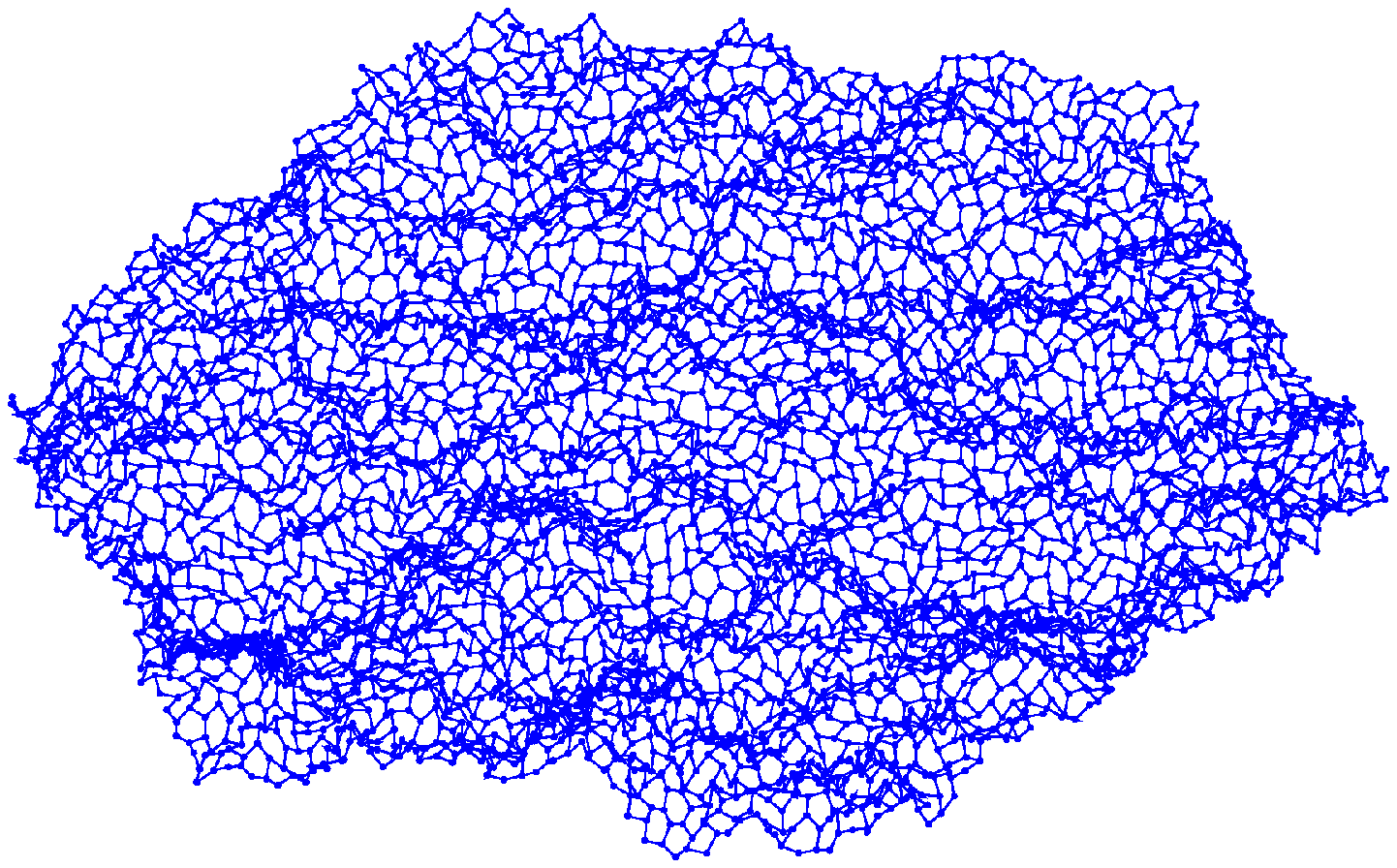}
\caption{[left panel] A  membrane with honeycomb structure that contains a total
of $N=54$ beads and has linear size $L=3$ ($L$ is the number of  hexagonal cells
on the edge of the membrane).   [right panel] A snapshot of a typical
conformation of an intact membrane with $L = 30$ containing $5400$ monomers
after equilibration with no external force applied. Typical wrinkles are seen
to form on  the surface.
\label{fig_ModelMembrane}}
\end{center}
\end{figure}

We study a coarse-grained model of honeycomb membrane embedded in
three-dimensional (3D) space. The membrane consists of $N$ spherical particles
(beads, monomers) of diameter $\sigma$ connected in a honey-comb lattice
structure whereby each monomer is bonded with three nearest-neighbors except for
the monomers on the membrane edges which have only two bonds (see
Fig.~\ref{fig_ModelMembrane} [left panel]). The total number of monomers $N$ in
such a membrane is $N = 6 L^2$ where by $L$ we denote the number of monomers (or
hexagon cells) on the edge of the membrane (i.e., $L$ characterizes the linear
size of the membrane). There are altogether $N_{bonds}=(3N-6L)/2$ bonds in the
membrane. In our studies we consider {\em symmetric} hexagonal membranes (i.e.,
{\em flakes}) so as to minimize possible effects due to the asymmetric of edges
or vortices at the membrane periphery.

For the analysis of our results we find it appropriate to divide the
two-dimensional membrane network so that all bonds fall into different subgroups
presented by concentric ``circles'' with consecutive numbers (see
Fig.~\ref{fig_ModelMembrane} [right panel]) proportional to their radial
distance from the membrane center. To {\em odd} circle numbers thus belong bonds
that are nearly tangential to the corresponding circle. {\em Even} circles
contain no encompass radially oriented bonds (shown to cross the circle in
Fig.~\ref{fig_ModelMembrane}). The total number of circles $C$ in a membrane of
linear size $L$ is found to be $C=(2L-1)$. We use this scheme of labeling the
groups of bonds that compose the membrane in order to represent our simulation
results in appropriate way which relates them to their relative proximity to
membrane's periphery.

\subsection{Potentials}

\begin{figure}[ht]
\begin{center}
\includegraphics[scale=0.40]{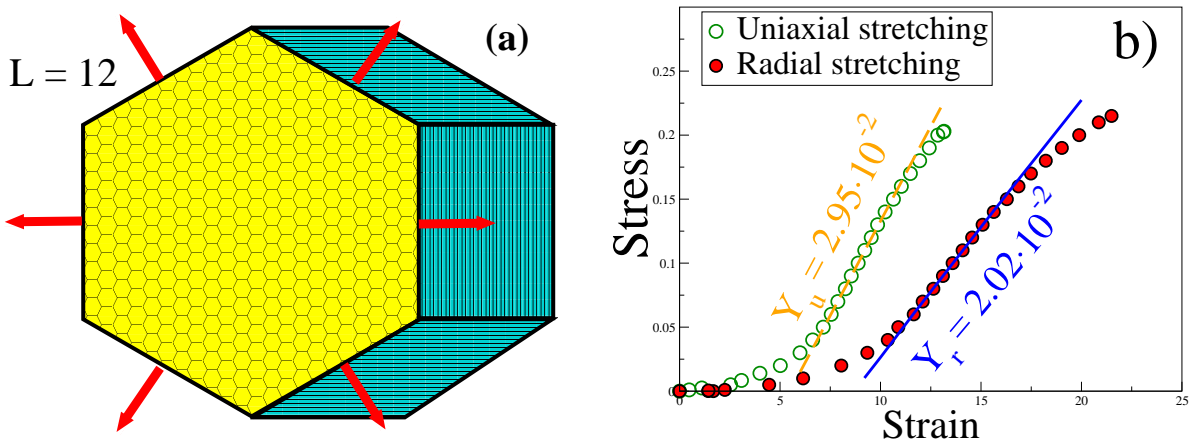}
\hspace{0.5cm}
\includegraphics[scale=0.29]{young.eps}
\caption{(a) A protective honeycomb network is spanned at the orifice of a prism
whose size may vary due to thermal expansion. Tensile forces acting on the
membrane periphery are indicated by red arrows. (b) Mean strain of a honeycomb
membrane of size $L = 10$ as a function of external tensile stress $f$ at $T =
0.01$ and $\gamma=0.25$. Depending on the way in which external force is
applied: (i) radial stretching - c.f. Fig.~\ref{fig_force_dir}f, or (ii)
uniaxial stretching - Fig.~\ref{fig_force_dir}b  the observed Young modulus is
$Y_r =2.02 \times 10^{-2}\; [k_BT/a^3]$  or $Y_u =2.95 \times 10^{-2}\;
[k_BT/a^3]$.
\label{fig_Young}}
\end{center}
\end{figure}

The nearest-neighbors in the membrane are connected to each other by breakable
anharmonic bonds described by a Morse potential,
\begin{equation}
U_{\text{M}}(r) = \epsilon_M \lbrace 1 - \exp[-\alpha(r-r_{\text{min}})] \rbrace
^{2}.
\label{eq_U_MORSE}
\end{equation}
 where $r$ is the distance between the monomers.
Here  $\alpha=1$ is a constant that determines the width of the potential well
(i.e., bond elasticity) and $r_{\text{min}} = 1$ is the equilibrium bond length.
The dissociation energy of a given bond, $\epsilon_M=1$, is measured in units of
$k_BT$ where $k_B$ denotes the Boltzmann constant and $T$ is the temperature.
The minimum of this potential occurs at $r=r_{\text{min}},\;~
U_{\text{Morse}}(r_{\text{min}}) = 0$. The maximal restoring force of the Morse
potential, $f_{\text{max}} = -dU_{\text{M}}/dr = \alpha \epsilon_M /2$, is
reached at the inflection point, $r = r_{\text{min}} + \alpha^{-1} \ln(2)
\approx 2.69$. This force $f_{\text{max}}$ determines the maximal tensile
strength of the membranes bonds. Since $U_{\text{M}}(0) \approx 2.95$, the Morse
potential, Eq.~(\ref{eq_U_MORSE}), is only weakly repulsive and beads could
partially penetrate one another at $r < r_{\text{min}}$. Therefore, in order to
allow properly for the excluded volume interactions between bonded monomers, we
take the bond potential as a sum of $U_{\text{M}}(r)$ and the so called
Weeks-Chandler-Anderson (WCA) potential, $U_{\text{WCA}}(r)$, (i.e., the shifted
and truncated repulsive branch of the Lennard-Jones potential),
\begin{eqnarray}
U_{\text{WCA}}(r) = \begin{cases}
4\epsilon \left[ \left( \frac{\sigma}{r} \right) ^{12} - \left( \frac{\sigma}{r}
\right) ^{6} \right] + \epsilon,  & \text{for}~~ r \leq 2^{1/6}\sigma \\
0,  & \text{for}~~ r > 2^{1/6}\sigma
\end{cases}
\label{eq_U_WCA}
\end{eqnarray}
with parameter $\epsilon=1$ and monomer diameter $\sigma=2^{-1/6} \approx 0.89$
so that the minimum of the WCA potential to coincides with the minimum of the
Morse potential. Thus, the length scale is set by the parameter $r_{\text{min}}
= 2^{1/6}\sigma = 1$. The nonbonded interactions between monomers are taken
into account by means of the WCA potential, Eq.~(\ref{eq_U_WCA}). Thus, the
nonbonded interactions in our model correspond to good solvent conditions
whereas the bonded interactions make the bonds  breakable when subject to
stretching.  External stretching force $f$ is applied to monomers at the
membrane rim in direction perpendicular to the respective edge -
Fig.~\ref{fig_Young}a.

Before we turn to the problem of membrane failure under constant tensile force,
we show here some typical elastic properties of the intact honeycomb
network sheet that is used in our computer experiments -  Fig.~\ref{fig_Young}.
In Fig.~\ref{fig_Young}b one can see an $S-$shaped variation of the stress -
strain relationship with initial significant elongation at vanishing stress
due to the straightening of the membrane wrinkles (ripples) that are
typical for an unperturbed membrane - cf. Fig.~\ref{fig_ModelMembrane}b. This
behavior is followed by a linear stress - strain elastic relationship where we
measure the Young modulus $Y_r =2.02 \times 10^{-2}\; [k_BT/a^3]$ (or $Y_u =2.95
\times 10^{-2}\; [k_BT/a^3]$), depending whether radial of uniaxial loading is
applied. Eventually,
for stronger stretching the elasticity  of the network decreases as the
anharmonicity of the bond potential comes into play.
Moreover Fig.~\ref{fig_Young}b indicates that the destructive strain of the
whole membrane is considerably less in the case of uniaxial stretching.

In our work we have tried to develop the model which should serve as generic
one for all kinds of 2D brittle-elastic networks with honeycomb orientation. We
have been anxious to emphasize the common features of failure in materials with
similar architecture but largely varying elasticity properties, e.g., from
$1000$~GPa graphene's Young modulus \cite{Aluru} compared to $4\times
10^{-3}$~GPa for spectrin \cite{Dao}. Putting the value of a Kuhn segment
($\sigma=1.44$~\AA) and taking the thermal energy $k_BT=4\times 10^{-21}$~J at
$T=300$~K, we get from our simulation a Young modulus $\sim0.03$~GPa which
% corresponds exactly to the measured value of polypropylene. 
is ranged between typical values for rubber-like materials 
$0.01\div0.1$.
Compared to {\it ab
initio} simulations of graphene with linear size $L=6$ which corresponds to
$216$ network
nodes \cite{Snook}, our objects are about an order of magnitude larger, $L=50$
and
$2500$ nodes, in units of elementary cells.

\subsection{MD algorithm}

As in our previous studies concerning scission kinetics of linear chains
\cite{Paturej,paturejjcp} and bottle-brushes \cite{milchev}  we use a Langevin
dynamics which describes the Brownian
motion of a set of interacting particles whereby the action of the solvent is
split into slowly evolving viscous (frictional) force and a rapidly fluctuating
stochastic (random) force. The Langevin equation of motion is the following:
\begin{equation}
m\overrightarrow{\dot{v}_i}(t) = \overrightarrow{F}_i(t) - m \gamma
\overrightarrow{v_i}(t) + \overrightarrow{R}_i(t)
\label{eq_Langevin}
\end{equation}
where $m$ denotes the mass of the particles which is set to $m=1$,
$\overrightarrow{v}_i$ is the velocity of particle $i$, $\overrightarrow{F}_i =
(\overrightarrow{F}_\textmd{M} + \overrightarrow{F}_\textmd{WCA})_i$ is the
conservative force which is a sum of all forces exerted on particle $i$ by other
particles in the system, $\gamma$ is the friction coefficient and
$\overrightarrow{R}_i$ is the three dimensional vector of random force acting on
particle $i$. The random force $\overrightarrow{R}_i$, which represents the
incessant collision of the monomers with the solvent molecules, satisfies the
fluctuation-dissipation theorem $\langle R_{i\alpha}(t) R_{j\beta}(t') \rangle =
2 \gamma k_B T \delta_{ij} \delta_{\alpha\beta} \delta(t-t')$ where the symbol
$\langle ... \rangle$ denotes an equilibrium average and the greek-letter
subscripts refer to the $x$, $y$ or $z$ components. The friction coefficient
$\gamma$ of the Langevin thermostat is set to $\gamma=0.25$. Our simulation was
performed in the weakly damped regime of $\gamma=0.25$ where effects of inertia
are important. This value of $\gamma$ is more or less standard in Langevin MD.
However, we carried out additional simulation in the strongly damped regime for
$\gamma=10$. No qualitative changes were discovered except an absolute overall
increase of the rupture times $\tau$ which is natural for a more viscous
environment. The integration step is $0.002$ time units (t.u.) and the time is
measured in units of $r_{\text{min}} \sqrt{m/\epsilon_M}$. We emphasize at this
point that in our coarse-grained modeling the solvent is taken into account only
implicitly. In this work the velocity-Verlet algorithm is used to integrate the
equations of motion.

Our MD simulations are carried out in the following order. First, we prepare an
equilibrated membrane conformation, starting with a fully flat configuration,
Fig.~\ref{fig_ModelMembrane}, where each bead in the network is separated by a
distance $r_{\text{min}}=1$ equal to the equilibrium separation of the bond
potential $(U_{\text{M}}+U_{\text{WCA}})$ [see Eq.~(\ref{eq_U_MORSE}) and
(\ref{eq_U_WCA})]. The external constant force is switched on from the very
beginning of the simulation. Then we start the simulation with this prepared
conformation and let the membrane equilibrate with the applied force in the heat
bath for sufficiently long time ($\approx 10^7$ t.u.) at a temperature that is
low enough so that the energy barrier for scission is high and the membrane
stays intact. This equilibration is done in order to prepare different starting
conformations for each simulation run. Once the equilibration is finished, the
temperature is raised to the working one and we let the membrane equilibrate at
this temperature for roughly $\sim 20$ t.u. We have checked that this time
interval is sufficient for equipartition and uniform distribution of temperature
to be established throughout the membrane sheet. Then the time is set to zero
and we continue the simulation with this well-equilibrated membrane conformation
checking for scission of the bonds.

We measure the elapsed time $\tau$ until the first bond rupture occurs and
repeat the above procedure for a large number of runs ($10^3\div10^4$), starting
each time with a new equilibrated conformation so as to sample the stochastic
nature of rupture and determine the mean $\langle \tau \rangle$ which we refer
to as the mean first breakage time. In the course of simulation we also
calculate properties such as the probability distribution of breaking bonds
regarding their position in the membrane (a rupture probability histogram), the
probability distribution function of the first breakage time $W(\tau)$ (i.e.,
the  MFBT probability distribution), the strain (extension) of the bonds with
respect to the consecutive circle number in the membrane, as well as other
quantities of interest. 

In separate runs each simulation  is terminated as soon as the
honeycomb sheet disintegrates into two separate parts whereby the time it takes
to ``rip-off'' the sheet is termed ``mean failure time $\langle \tau_r
\rangle$ and measured. In order to monitor the propagation of cracks, we
perform also individual runs labeling breaking bonds in succession and
reconstructing the crack trajectory which is a laborious and rather involved
problem.

\subsection{Rupture criterion}

An important aspect of our simulation is the recombination (self-healing) of
broken bonds. The constant stretching force acting on the monomers at the
membrane edges creates a well-defined activation barrier for bond scission.
Direct analysis of the one-bond potential with external force, $U_{\text{M}}(r)
- fr$ indicates that the positions of the (metastable) minimum $r_{-}$ and of
the barrier (or hump) $r_{+}$ are given by \cite{ghosh}
\begin{eqnarray}
 r_{- , +} = \dfrac{1}{a} \: \ln \Biggl[ \dfrac{2}{1 \pm \sqrt{1 -
{\tilde f}}}\Biggr]
\label{Min_Max}
\end{eqnarray}
where the dimensionless force ${\tilde f} = 2 f/a \epsilon_M$. For the range of
tensile forces used in the present study one has typically $r_+ \approx
3r_{\text{min}}$. The activation energy (barrier height) for single bond
scission  is  itself given by \cite{ghosh}
\begin{eqnarray}
 E_{b} = U (r_{+}) - U (r_{-}) = \epsilon_M \left\lbrace \sqrt{1 - {\tilde f}} +
\dfrac{{\tilde f}}{2}  \:  \ln \Biggl[ \dfrac{1 - \sqrt{1 - {\tilde f}}}{1
+  \sqrt{1 - {\tilde f}}}\Biggr]\right\rbrace
\label{eq_Barrier}
\end{eqnarray}
One can easily verify that $E_b$ decreases with ${\tilde f}$. Since a bond may
get stretched beyond the energy barrier and nonetheless shrink back again, i.e.
recombine, in our numeric experiments we use a sufficiently large value for
critical extension of the bonds, $r_h=5r_{\text{min}}$, which is defined as a
threshold to a broken state of the bond. This conventions is based on our checks
that the probability for recombination (self-healing) of bonds, stretched beyond
$r_h$, is vanishingly small, as demonstrated below. In our model we deal
 with $E_b/(k_BT)=20$ which at $300$~K and bond length $r_{min}=0.14$~nm
corresponds to ultimate tensile stress $\sim 0.6$~GPa. This is a reasonable
value for our membrane which is considerably softer than graphene with
$\sim100$~GPa \cite{Aluru} and is ranged between typical values for rubber
materials $~0.03\div 14$~GPa.

\section{MD-results} \label{sec_MD}

\begin{figure}[ht]
\begin{center}
\includegraphics[scale=0.23, angle=0]{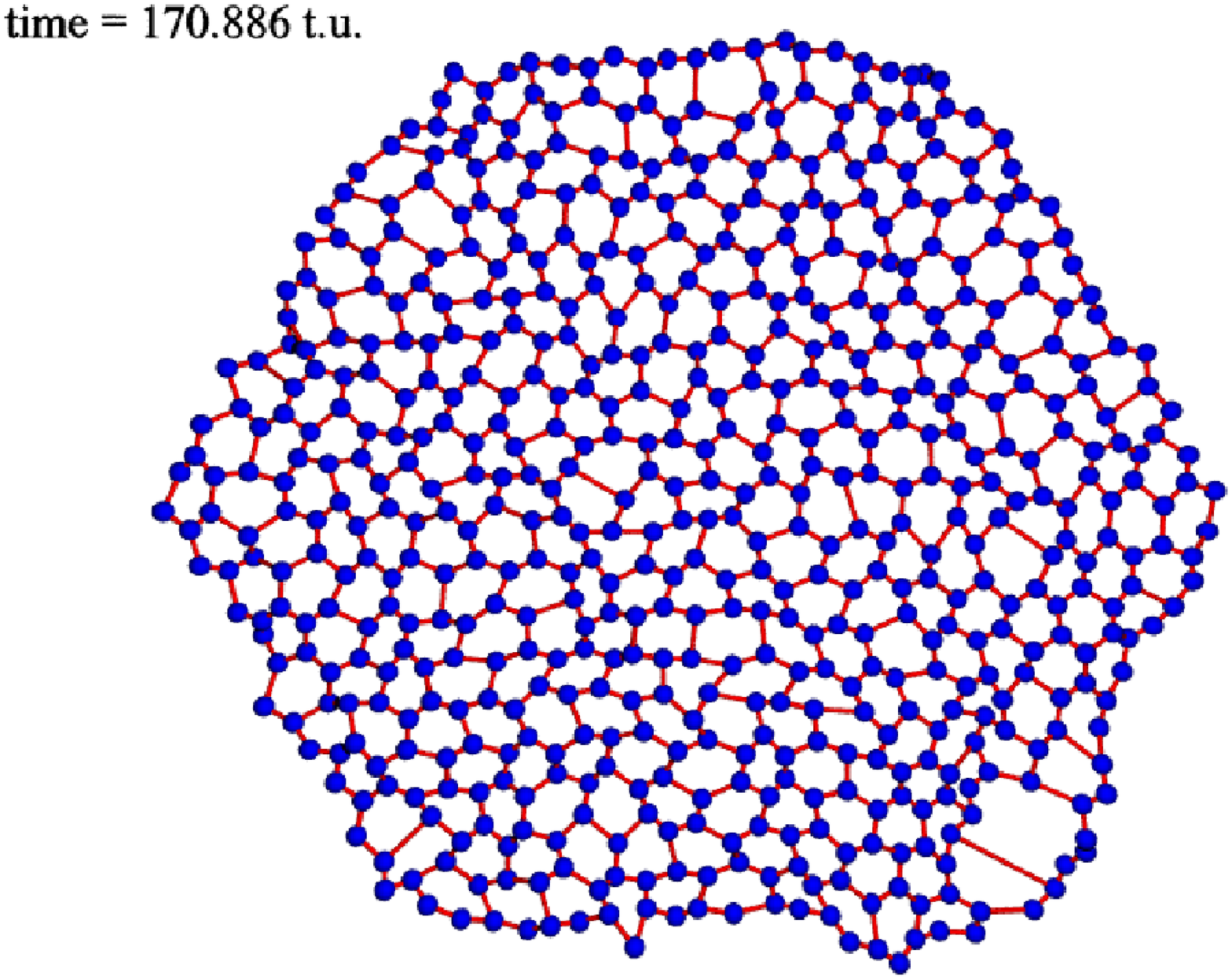}
\includegraphics[scale=0.23, angle=0]{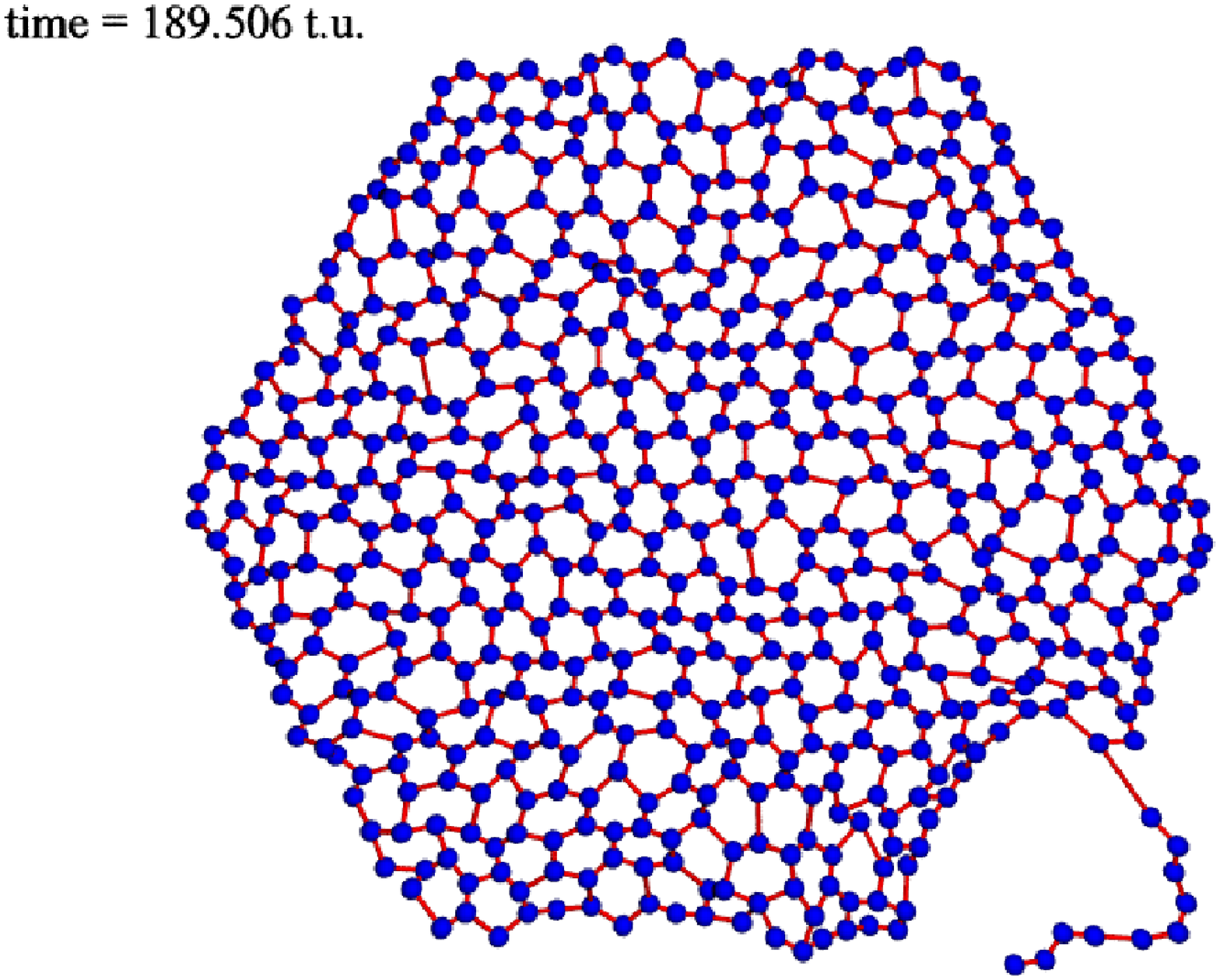}
\includegraphics[scale=0.23, angle=0]{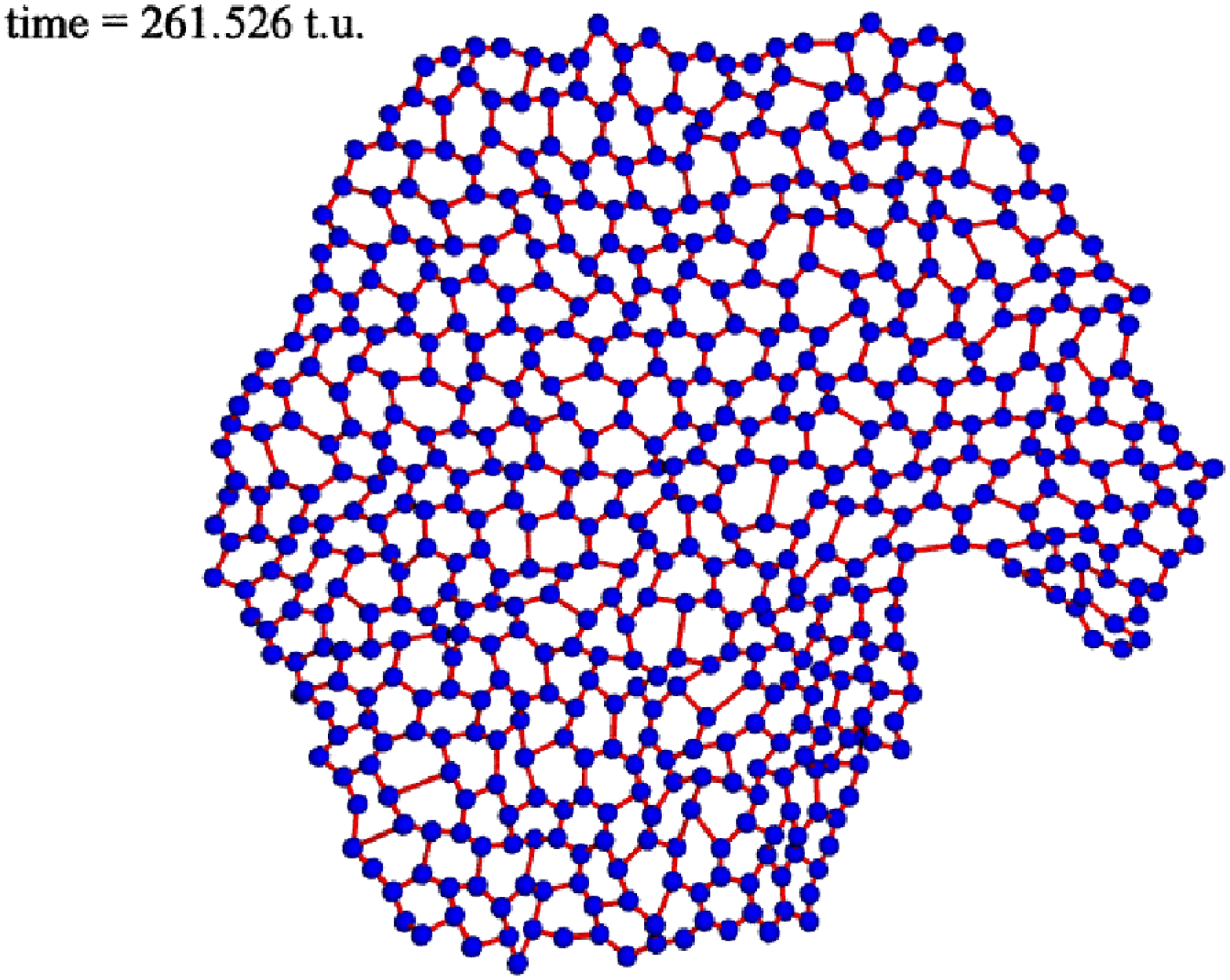}
\includegraphics[scale=0.23, angle=0]{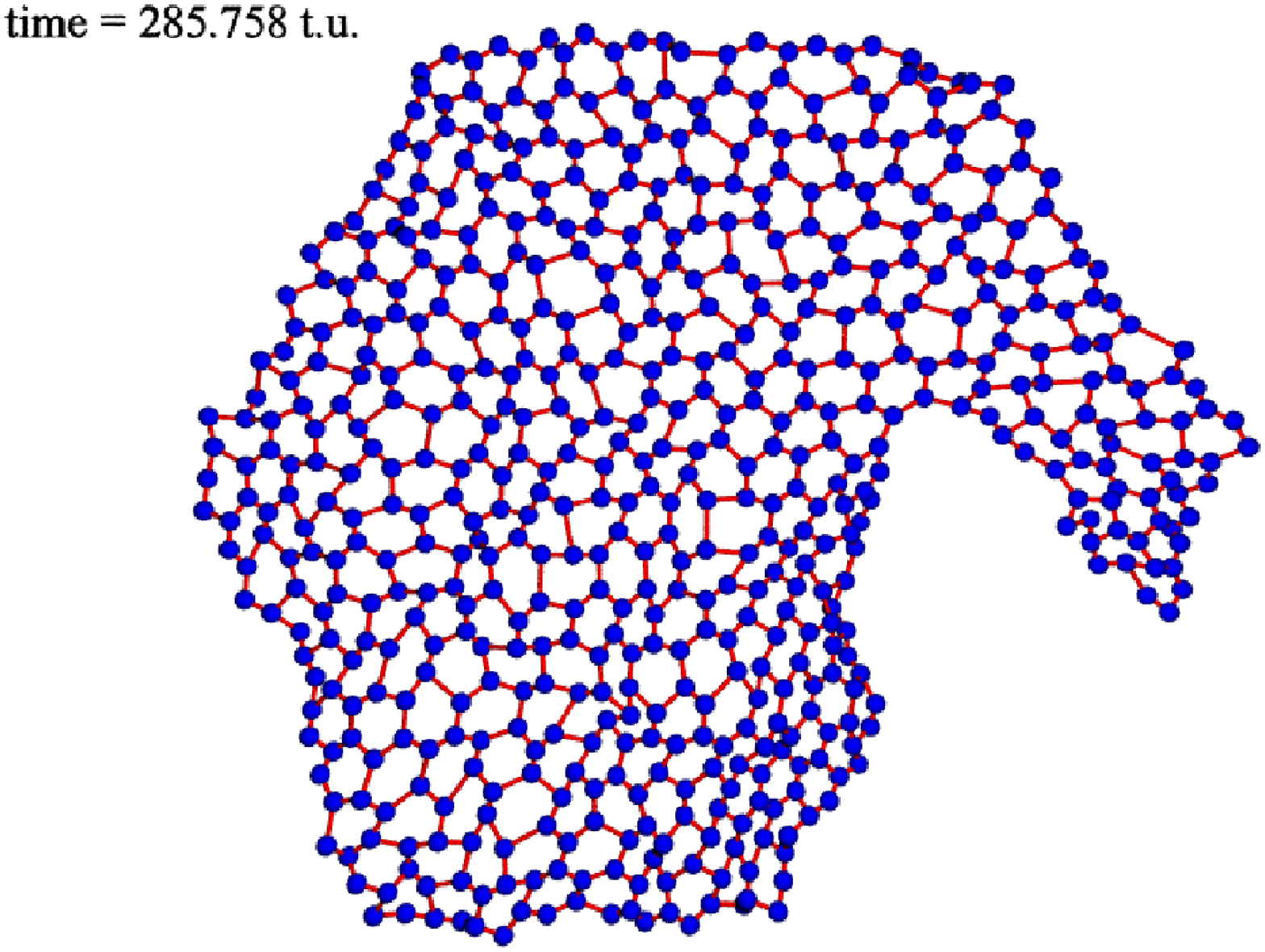}
\includegraphics[scale=0.23, angle=0]{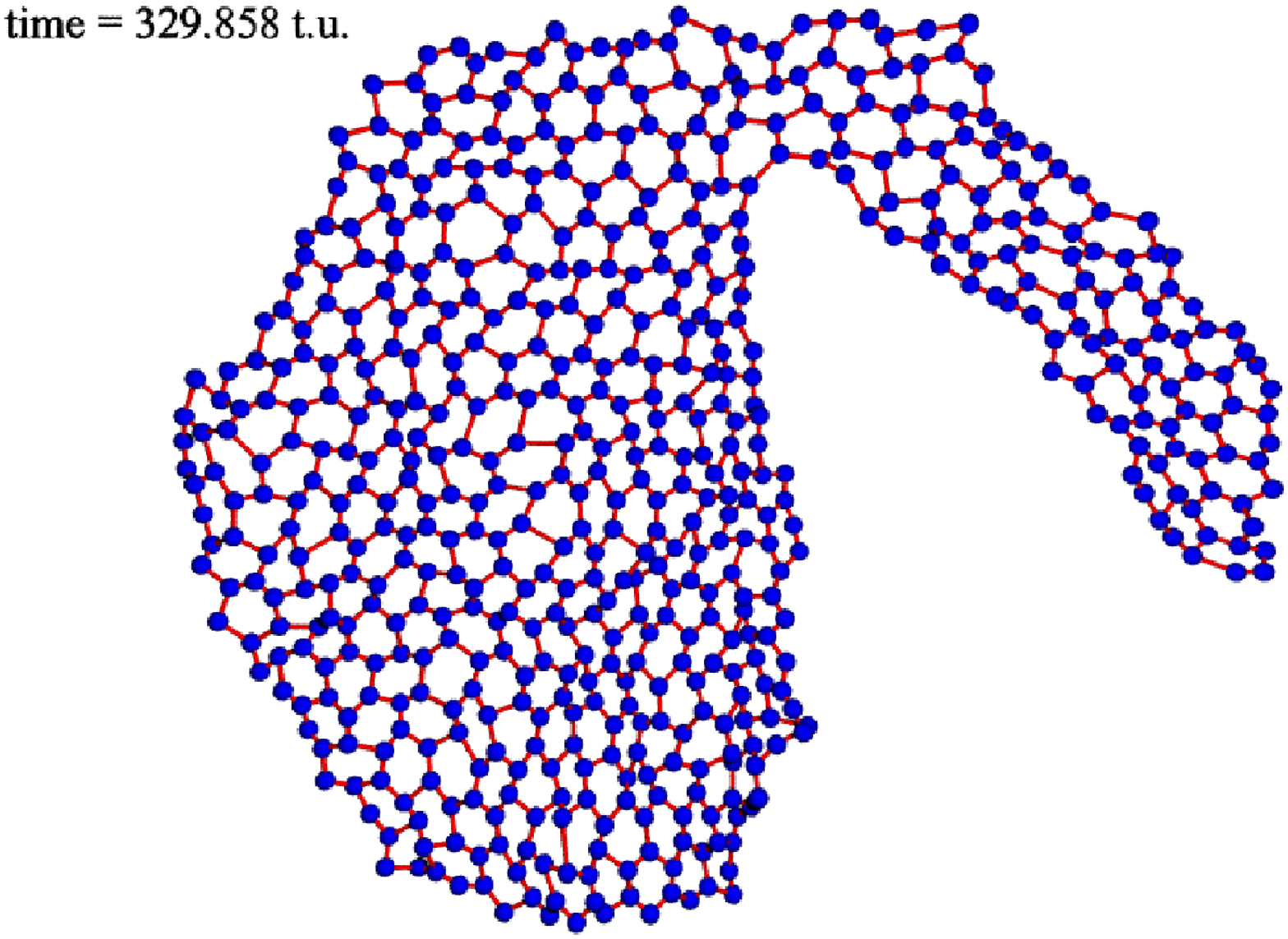}
\includegraphics[scale=0.23, angle=0]{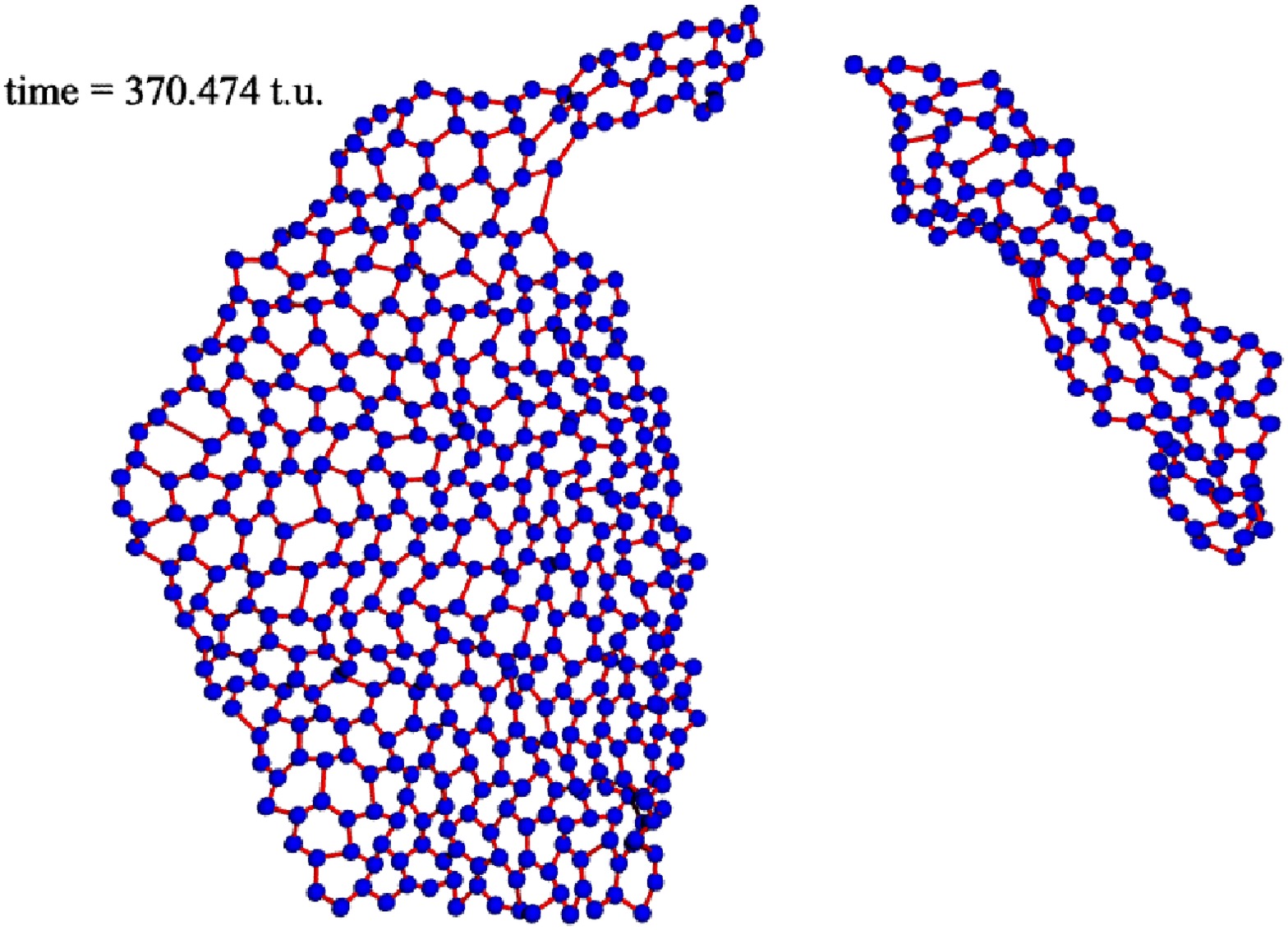}
\caption{Snapshots illustrate the process of bond breakage (crack generation) in
different time moments for a membrane with $N=600$ particles subject to external
tensional force $f=0.15$ at $T=0.05$ and $\gamma=0.25$. The force is applied to
periphery monomers
only and stretches the network perpendicular to its original edges.
\label{fig_snapshots_break}}
\end{center}
\end{figure}

We examine the scission of bonds between neighboring nodes in the network sheet
with honeycomb topology, assuming thermal activation as a driving mechanism in
agreement with early experimental work by Brenner \cite{Brenner} and Zhurkov
\cite{Zhurkov}. In Fig.~\ref{fig_snapshots_break} we show a series of
representative snapshots of a membrane of size $L = 10$ with $N=600$ monomers
taken at different time moments during the process of decomposition. Typically,
the first bonds that break are observed to belong to the last (even) most remote
circle as, for example, at $t \approx 171 t.u.$ in
Fig.~\ref{fig_snapshots_break}. As mentioned above, these are the radially
oriented bonds which belong to concentric circles of even number. Gradually a
line of edge beads is then severed from the rest of the membrane and a crack is
formed which propagates into the bulk until eventually a piece of the network
sheet is ripped off, as in Fig.~\ref{fig_snapshots_break} at $t \approx 370
t.u.$ As we shall see below, this mechanism of membrane failure, whereby an
initial crack is formed parallel to the edge monomers, yet perpendicular to the
tensile force, dominates largely the process of disintegration under constant
tensile force. The process is, therefore, mainly described by two characteristic
times, $\tau$ and $\tau_r$, which mark the occurrence of the first scission of a
bond (MFBT) and that of the eventual breakdown of the flake into two distinct
parts.

\subsection{Bond recombination}\label{subsec_rec}

\begin{figure}[ht]
\begin{center}
\includegraphics[scale=0.4]{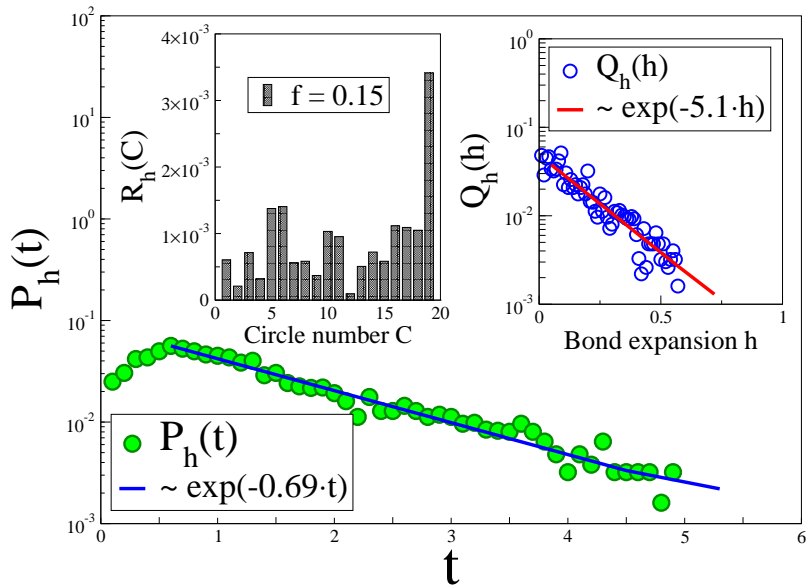}
\caption{Probability distribution $P_h(t)$ of maximal times (full circles), and
$Q_h(h)$ of maximal bond lengths $h$ (circles, right panel of inset) before a
recombination event in the stretched membrane with $N=600$, $T=0.1$,
$\gamma=0.25$ takes places. The exponential tail of $P_h(t)$ is fitted by blue
line. The exponential decay of $Q_h(h)$ is given by red line. The left panel of
the inset shows the healing probability $R_h$ vs.~circle number $C$. The healing
events under applied stress occur roughly 10 times less frequently than for $f =
0$. \label{fig_healing}}
\end{center}
\end{figure}

As mentioned in Section \ref{sec_MD}, throughout in our studies of the brittle
sheet breakdown we use a threshold for critical bond stretching (rupture
criterion) $r_h = 5 r_{min}$.  In the right inset of Fig.~\ref{fig_healing} we
display the function
$Q_h(h)$, which represents the probability distribution of  bond stretching $h$
beyond the hump position $r_+$,  given that a subsequent recombination has
taken place. To this end one monitors for $10^4$ integration steps the length
of each bond once the bond expands beyond $r_+$ and stores its maximal
expansion, $h$, provided such a bond contracts again to $r < r_+$.  Then
$Q_h(h)$ is computed as the fraction of extensions to $h$ over the total number
of recombination events. For each bond recombination one measures also the
distribution of the respective self-healing times, $P_h(t)$, which is shown in
Fig.~\ref{fig_healing} too. Both distributions are characterized
by exponentially fast decaying tails, indicating that successful recombinations
are possible after very short
time interval $\approx 1.3 t.u.$, and the possible stretching of a bond in such
cases is minimal - about $0.19 \div 0.5 $ beyond the energy barrier position at
$r_+ \approx 2.96$, that is, significantly shorter than $r_h \approx 5$. We also
find that recombination of bonds takes place seldom (roughly 1.5\% over
$5\cdot10^4$ runs of average length $\approx 437 t.u.$ for a membrane composed
of $N=600$ beads). Yet as indicated below, allowing for self-healing events may
significantly change the observed kinetics of membrane destruction. The left 
inset in Fig.~\ref{fig_healing} indicates that self-healing of bonds happens
most frequently at the membrane periphery, $C=19$, where bond stretching occurs
most frequently.

\subsection{Mean First Breakage Time}\label{subsec_MFBT}

These conclusions, based on visual evidence from snapshots taken in the course
\begin{figure}[ht]
\begin{center}
\includegraphics[scale=0.18]{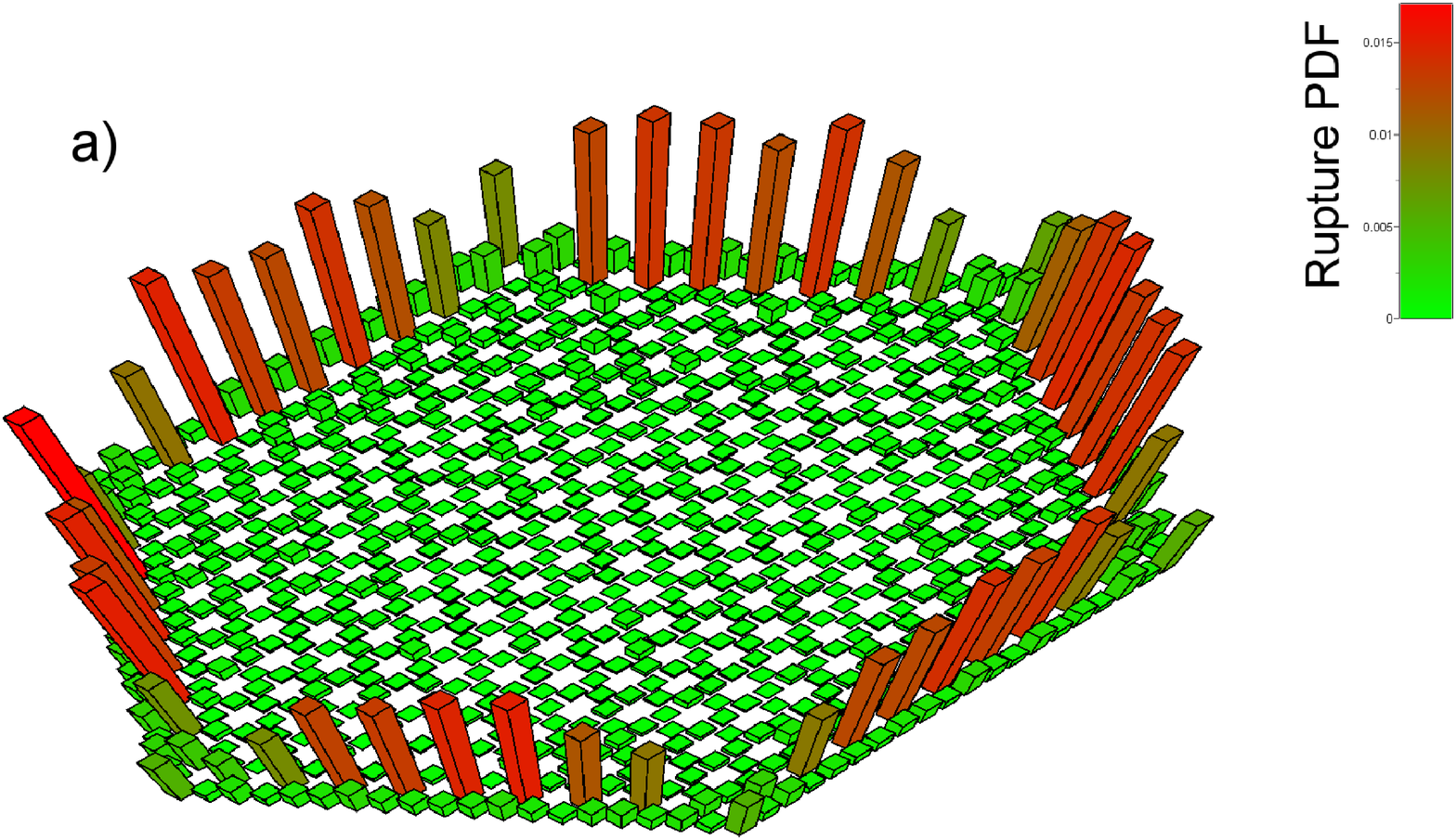}
\hspace{0.5cm}
\includegraphics[scale=0.23]{circ.eps}
\caption{(a) Rupture probability histogram of flexible hexagonal membrane 
subjected to external tensile stress $f=0.15$. 
(b) Scission probability histogram vs consecutive circle number for membrane
pulled with $f=0.125$ displayed for different rupture thresholds $r_h$ as
indicated.
Real rupture events ($r_h=5.0$) are concentrated at the periphery
whereas fictitious ones ($r_h=3.1$) are distributed more uniformly all over the
membrane.
Here $N=600$, $T=0.05$ and $\gamma=0.25$.
 \label{fig_histo_rate}}
\end{center}
\end{figure}
of membrane decomposition, are corroborated in Fig.~\ref{fig_histo_rate}a where
we show the probability distribution of a first rupture for{\em all} bonds in
the honeycomb membrane flake as a 3D plot. It is seen that the scission rate is
localized in the outer-most circle of radial bonds whereas bonds in the inner
part of the membrane practically hardly break. Note that this is not a trivial
effect since tension is 
distributed uniformly over all bonds in the equilibrated membrane so there is
no additional propagation of the tension front from the rim towards the center.
Fig.~\ref{fig_histo_rate}b also indicates a qualitative
change in the rupture PDF when self-healing is not allowed for (by reducing the
threshold position to that of the energy barrier - $r_h=3.1$) in contrast to
results where self-healing was fully taken into account - $r_h=5$.  Moreover, a
closer inspection the new Fig.~\ref{fig_histo_rate}b indicates that scission of
bonds with no self-healing takes place almost uniformly throughout the membrane
while with self-healing it is concentrated only at the membrane  periphery.

\begin{figure}[ht]
\begin{center}
\includegraphics[width=0.4\textwidth]{l2.eps}
\hspace{0.5cm}
\includegraphics[totalheight = 0.55\textwidth, origin = br, angle=-90]
{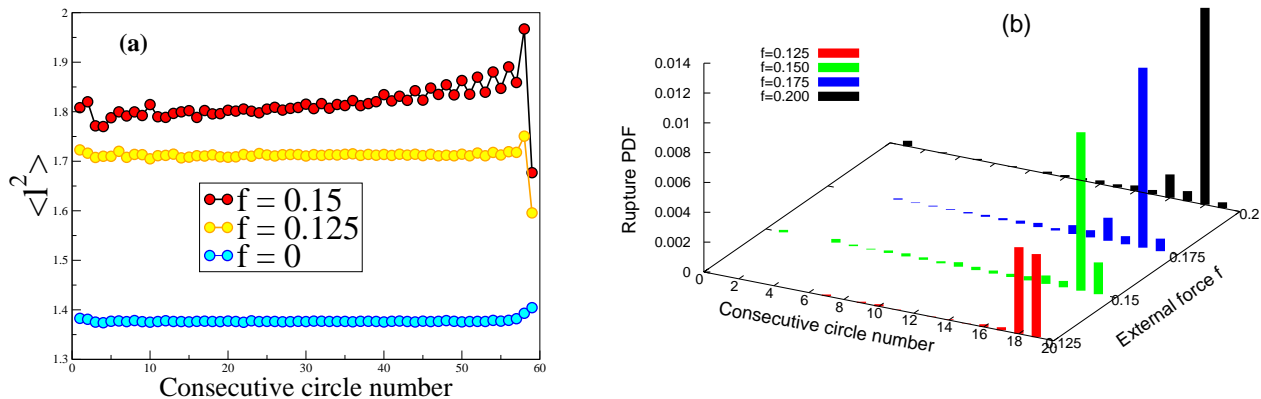}
\caption{(a) Variation of the mean squared bond length $\langle l^2 \rangle$
with consecutive circle number in a membrane with $N=5400$ beads subjected to
different strengths of the external force $f$ (as indicated in the legend).
(b) Probability
distribution of the first bond scission event vs circle number in a membrane
with $N=600$ beads at different strengths of the external force $f$ as
indicated. For a force $f \leq 0.15$ the bonds from the last two outer circles
($\#18$ and $\#19$) in the membrane have the highest rupture probability. With
increased strength of the pulling force $f\geq 0.175$ the bonds which are
located in the circles $\#18$ and $\#16$ attain the highest rupture
probability.   
Parameters of a heat bath are $T=0.05$ and $\gamma=0.25$.
\label{fig_rate_strain}}
\end{center}
\end{figure}

One can try to relate this finding to the distribution of strain within the
network as shown in Fig.~\ref{fig_rate_strain}a and sampled for several
strengths of the
external stretching force $f$. In the case of strongest pulling, $f = 0.15$, the
variation of the mean-squared bond length $\langle l^2 \rangle$ with distance
from the membrane center (i.e., with consecutive circle number $C$) displays a
well expressed saw-tooth behavior whereby the peaks correspond to bonds with
radial rather than tangential orientation (odd $C$). The alternation of strongly
/ weakly stretched bonds
modulates the overall gradual increase of the mean bond length with
growing distance from the center. Evidently, the amplitude of the mean-squared
bond length attains a pronounced maximum on the last circle of radially oriented
network bonds. This distribution of strain is found to persist down to
vanishing tensile force $f = 0$ -  Fig.~\ref{fig_rate_strain}a.
The distribution of first scission events is clearly seen in
Fig.~\ref{fig_rate_strain}b where we show it for several strengths of $f$.
Evidently, with growing value of $f$ bonds happen to break also deeper inside
the membrane although such events remain much less probable.

\begin{figure}[ht]
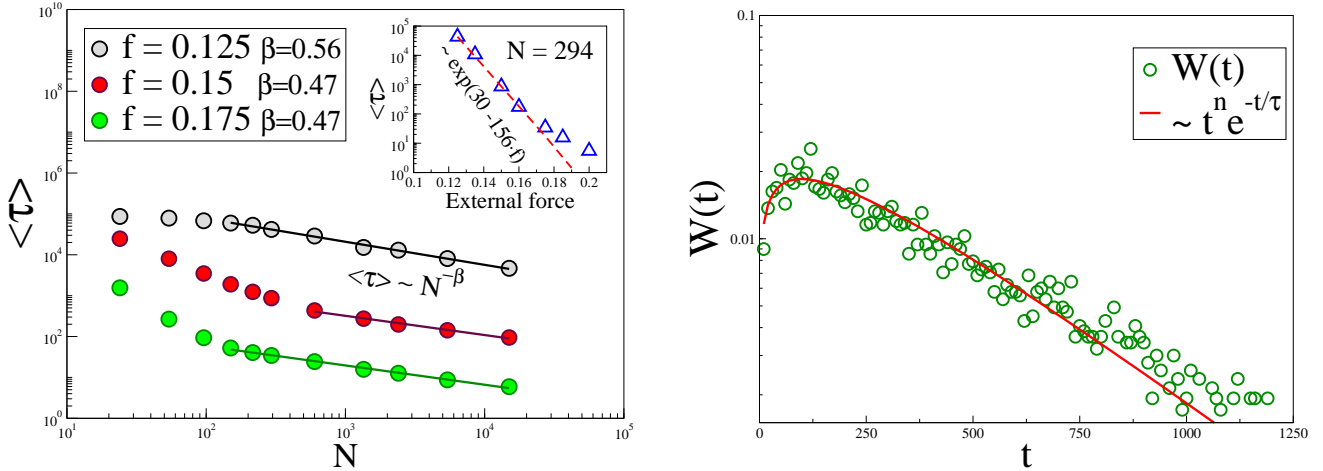

\vspace{0.7cm}
\begin{center}
\includegraphics[scale=0.31]{tauvsNforce.eps}
\hspace{0.5cm}
\includegraphics[scale=0.32]{Wtau.eps}
\caption{(a) Mean first breakage time $\langle \tau \rangle$ vs.~number of
particles $N$ in the membrane pulled with different tensile stress $f$ as
indicated. Symbols represent simulation data whereas solid lines stand for
fitting functions $\langle \tau \rangle \sim N^{-\beta}$. The inset shows
force-dependent $\langle \tau \rangle$ for a membrane composed of $N=294$ beads.
(b) MFBT probability
distribution $W(t)$ for the first scission of a bond in a flake with $N=600$
particles at stress $f = 0.15$. Symbols denote result of simulation and full
line stands for the fitting function $W(t) \propto t^n\exp(-t/\tau)$ with $n =
1/3$ and $\tau =  291.85 t.u.$
Parameters of heat bath are $T=0.05$ and $\gamma=0.25$.
\label{fig_MFBT}}
\end{center}
\end{figure}

The variation of the MFBT $\tau$ with system size $N$
(i.e., with the number of monomers in the membrane $N = 6 L^2$ where $L$ denotes
the linear size of the flake) is shown in Fig.~\ref{fig_MFBT}.
For sufficiently large membranes one observes a power law decline of the MFBT,
$\tau \propto N^{-\beta}$ with an exponent $\beta \approx 0.5\pm0.03$ for
the tensile
forces studied.. If thermally activated bonds break independently from one
another and entirely at random, then the MFBT $\tau$ measures the interval
before {\em any} of the available intact bonds undergoes scission, that is,
either the first bond breaks, or the second one, and so on which, at constant
rate of scission, would reduce the MFBT $\tau \propto 1/N$ as observed for
instance in the case of thermal degradation of a linear polymer chains
\cite{paturejjcp}. A more comprehensively this simple result can be derived by
means of the classical theory of Weibull. In the present system of a honeycomb
membrane the bonds that undergo rupture are nearly all located at the rim of the
flake and their number is proportional to $L$ so that with $\beta \approx 0.5$,
cf. Fig.~\ref{fig_MFBT} and $N \propto L^2$,  one obtains eventually the
important result $\tau \propto 1 / L$. This observation is in agreement with
recent results of Grant et al. \cite{Grant} who studied the nucleation of cracks
in a brittle $2D$-sheet. We should like to point out at this place that without
self-healing, c.f., Fig.~\ref{fig_histo_rate}b,   rupture time goes as
$\langle\tau\rangle\propto N^{-\beta}$, with $\beta\approx1$ (not plotted here)
in contrast to the observed exponent $\beta\approx 0.5$.

One should mention here an interesting analogy between the observed power-law
dependence of the MFBT time on system size and the power-law decrease of
life-time with system size in thermally activated breakdown of fiber bundles
\cite{kun1,kun2}. While both in our honey-comb network as well as in the Fiber
Bundle Model (FBM) the failure mechanism is related to redistribution of load
on neighboring bonds (fibres) upon single rupture, bonds in our membrane are
subject to a single scission threshold whereas in the FBM there is a random
distribution of tensile strengths. As a result, one finds a single value of
$\beta \approx 0.5$ and an Arhenian dependence of $\tau$ on temperature $T$ in
our elastic-brittle honey-comb network (see below) while the exponent
$\beta$ depends on the external load $f$ and on $T$ giving rise to a
non-Arhenian $\tau$ vs.~$T$ relationship. 

 Note  that the decline of MFBT $\tau$ in a topologically
connected  brittle system is by no means a trivial one. In a recent study
\cite{Paturej}, using Molecular Dynamics (MD) simulation of a single anharmonic
polymer chain  subject to constant external tensile force, we found a rather
complex interplay between the polymer chain dynamics and the resulting bond
rupture probability distribution along the chain backbone. In a breakable chain
(rather than 2D network) it was observed that the corresponding power $\beta
\rightarrow 0$ as $N \rightarrow \infty$. A major role in this was attributed to
nonlinear excitations as the possible origin for the observed increasing
insensitivity of rupture time with respect to polymer length as the pulling
force grows. One may thus conclude that nonlinear effects in bond scission are
suppressed in 2D honeycomb networks.

One can also see from the inset in Fig.~\ref{fig_MFBT}a that the MFBT $\tau$
decreases rapidly with growing stress $f$, that is, the energy barrier for
rupture declines with $f$ in agreement with Eq.~(\ref{eq_Barrier}) and
Zhurkov's experiments \cite{Zhurkov}. The probability distribution of MFBT
$W(t)$ is shown in Fig.~\ref{fig_MFBT}b. It is well described by a Poisson
probability distribution function $W(t) = 5.57\cdot 10^{-3}
t^{1/3}\exp(-t/291.85)$.

\subsection{Cracks and Mean Failure Time}
\label{subsec_cracks}
\begin{figure}[ht]
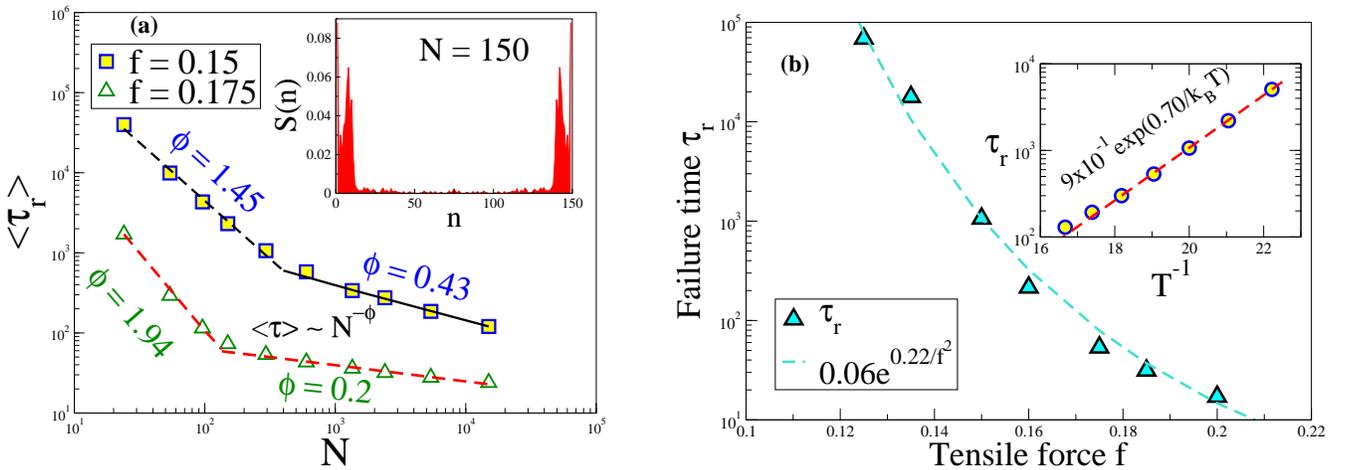

\begin{center}
\includegraphics[scale=0.31]{ripoff.eps}
\hspace{0.7cm}
\includegraphics[scale=0.35]{ripoff_fit.eps}
 \caption{(a) Mean failure time $\langle \tau_r \rangle$ (time needed to split
membrane into two pieces) vs.~number of particles in the membrane for two values
of the external pulling force $f$ at $T=0.05$ and $\gamma=0.25$.  Symbols denote
simulation results and solid
line represents power law fitting function $\langle\tau_r\rangle \sim
N^{-\phi}$. The inset shows PDF of number of particles in the moment of
splitting for a membrane composed of $N = 150$ beads. (b) Failure time $\langle
\tau_r \rangle$ vs $f$ in the case of $N = 294$. \label{fig_tau_rip_off}}
\end{center}
\end{figure}
The variation of $\tau_r$,
the mean failure time of the membrane with system
size $N$, shown in Fig.~\ref{fig_tau_rip_off}a, 
displays also a power-law dependence on system size $N$,
$\langle\tau_r\rangle\propto N^{-\phi}$, whereby
$\phi$ undergoes a cross-over to a lower value beyond roughly $N>300$.
%behaves qualitatively similar to
%the MFBT $\tau$ vs $N$ relationship. 
However, $\tau_r$ has different physical
meaning. Following Pomeau \cite{Pomeau}, the failure time can be approximately
identified with the nucleation of a crack of critical size $l_c$ given by
Griffith's critical condition \cite{Griffith,Rabinovich} assuming that crack
propagation is much faster than the nucleation time. For a $2D$-geometry
consisting of a flat brittle sheet with a crack perpendicular to the direction
of stress, the potential energy per unit thickness of the sheet reads $U = -
\frac{\pi l^2 f^2}{4Y} + 2\varepsilon l + U_0$ where $Y$ is the Young modulus,
$\varepsilon$ is the surface energy needed to form a crack of length $l$, and
$U_0$ is the elastic energy in the absence of stress ($f = 0$). This energy
reaches a maximum for a critical crack length $l_c = \frac{4\varepsilon Y}{\pi
f^2}$ beyond which no stable state exists except the separation of the sheet
into two broken pieces. Thus, with a crack nucleation barrier $\Delta U =
\frac{4\varepsilon ^2 Y}{\pi f^2}$ (in $3D$ $\Delta U \propto f^{-4)}$), the
failure (rip-off) time $\tau_r = \tau_0 \exp (\Delta U_0 / k_BT)$ as found in
experiments with bidimensional micro crystals by Pauchard and Meunier
\cite{Pauchard} and in gels by Bonn et al. \cite{Bonn}. In
Fig.~\ref{fig_tau_rip_off}b we present the variation of $\tau_r$ for membrane
failure with stress $f$ in good agreement with the expected relationship $\Delta
U \propto f^{-2}$. In addition, we show the variation of $\tau_r$ with
temperature, see inset in Fig.~\ref{fig_tau_rip_off}b, which is found to follow
a well expressed Arhenian relationship with inverse temperature, in agreement
with earlier studies \cite{Grant,Rabinovich}.

The end of the sheet rupturing process is marked as a rule by disintegration
into two pieces of different size so it is interesting to asses the size
distribution of such fragments upon failure. In the inset in
Fig.~\ref{fig_tau_rip_off}a we show a probability distribution $S(n)$ of the
sizes of of both fragments upon membrane rip-off. In a membrane composed of $N$
beads one observes a sharp bimodal distribution with narrow peaks at sizes $N_1
\approx 10$ and $N_2 \approx 140$. Evidently, for the adopted nearly radial
direction - cf. Fig.~\ref{fig_Young}a - of the applied tensile force one always
finds a pair of one small and another very large fragment.

\begin{figure}[ht]
\begin{center}
\includegraphics[scale=0.27]{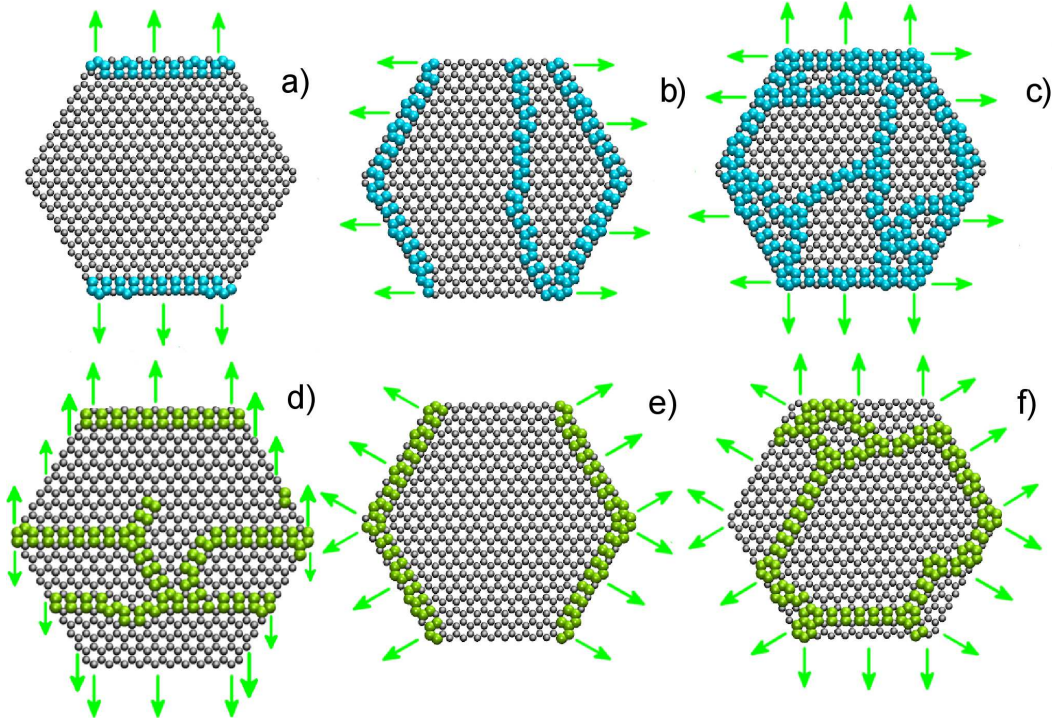}
\caption{ Overview of observed cracks in a honeycomb membrane composed of
$N=600$ particles for different orientation of the applied external pulling
force. Green arrows show indicate the orientation of the applied force
($f=0.15$): (a), (b), (d) - uniaxial, (c) - biaxial, (e), (f) - slanted.
Parameters of a heat bath are $T=0.05$ and $\gamma=0.25$. The typical cracks
are marked in color on the geometrically undistorted arrangement of
network nodes for better visibility. \label{fig_force_dir}}
\end{center}
\end{figure}
One can readily verify from the typical topology of the observed cracks in the
\begin{figure}[ht]
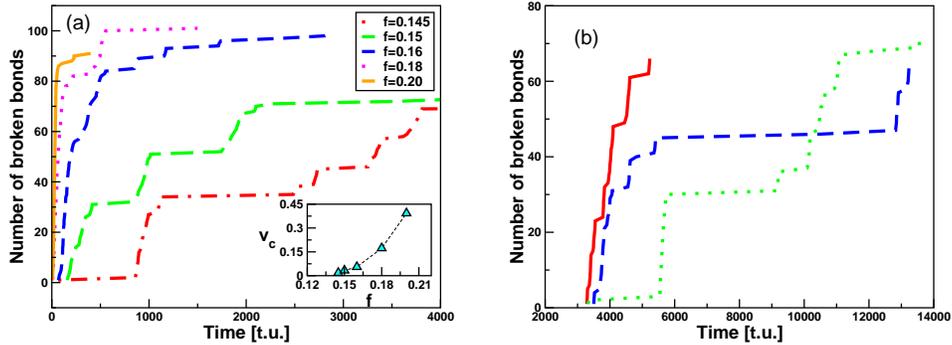

\begin{center}
\vspace{0.7cm}
\includegraphics[scale=0.31, angle=0]{V_crack_L10_forces.eps}
\hspace{0.5cm}
\includegraphics[scale=0.31, angle=0]{V_crack_L10_f0.14_1.2.3.eps}
\caption{(a) Crack propagation velocity (number of broken bonds per unit time)
for a membrane with $N=600$ beads at different strength of the external force
$f$ as indicated. (b) Three different realizations of cracks at applied force
$f=0.14$. The inset shows a variation of the mean crack propagation velocity
with $f$. Here $T=0.05$ and $\gamma=0.25$. \label{fig_cracks}}
\end{center}
\end{figure}
membrane, presented in Fig.~\ref{fig_force_dir}, that  (i) cracks emerge as a
rule perpendicular to the direction of applied stress, and (ii) it is almost
always the first row of nodes to which the tensile force is immediately applied
that gets ripped off upon failure. Cracks that break the network sheet in the
middle occur very seldom, in compliance with the sampled distribution of
fragment sizes, $S(n)$ in the inset of Fig.~\ref{fig_tau_rip_off}a. One would,
therefore, predict a breakup of a protective cover spanned on the orifice of
tube like the one shown in Fig.~\ref{fig_force_dir} to proceed immediately at
the fixed orbicular boundary where the tensile force applies to the network.
It is interesting to note that the geometry of cracks in the membrane shown in
Fig.~\ref{fig_force_dir} appears very similar to the one observed in drying
induced cracking of thin layers of materials subject to structural disorder
\cite{kun3}.

The emerging cracks are expected to propagate with speed that increases as the
strength of the external force is increased as the inset in
Fig.~\ref{fig_cracks}a indicates. In fact, in Fig.~\ref{fig_cracks}a one
observes typical curves comprising a series of short intervals with steep growth
of the number of broken bonds per unit time and longer horizontal 'terraces'
preceding the nucleation of a new crack. Even though the data, presented in
Fig.~\ref{fig_cracks}a, is not averaged over many realizations, and, as
Fig.~\ref{fig_cracks}b suggests, individual realizations of propagating cracks
may strongly differ even at the same stress $f$, a general increase of the
propagation velocity with growing external force $f$ - see inset - can be
unambiguously detected, in agreement with earlier observations \cite{Holland}.

For our model membrane with computed Young modulus $Y\approx 0.02$ we get for
the Rayleigh wave speed $c_R\approx 0.14$. Thus for most of the applied tensile
stress values we observe crack propagation at speed less than $c_R$ - inset in
Fig.~\ref{fig_cracks}b. As argued by \cite{Fineberg} propagation speed cannot
exceed $c_R$ because crack splits off into multiple cracks before reaching
$c_R$. In contrast, Abraham and Gao in Ref.~\cite{Abraham} have reported on
cracks that can travel faster than the Rayleigh speed. Thus, our rough estimates
(inset in Fig.~\ref{fig_cracks}) agree well with data from literature.
Converting our results to proper metric units, with bond length $\sigma\approx
0.144$~nm and energy $\approx 20 k_BT$ which yields 1 MD t.u.~$\approx
10^{-12}$~s,
we estimate the typical crack propagation speed $v_c\approx 50$~m/s.
Note that
 mean crack speed  for natural latex rubber was given as $56$~m/s
\cite{petersan}.

\section{Summary}\label{sec_summary}

In the present work we have studied the bond rupture and ensuing fracture of a
honeycomb brittle membrane subject to uniform radially applied external
stretching forces for different values of  force $f$, temperature $T$, and
membrane size $N$. The most important conclusions that can be drawn from our
Molecular Dynamics simulation can be summarized as follows:

\begin{itemize}
 \item bonds scission in hexagonal 2D sheets with honeycomb structure of the
underlying network under subjected to external pulling perpendicular to flake's
edges take place overwhelmingly at the sheet periphery
 \item The Mean First Breakage Time of breaking bonds depends on membrane size
$N$ as a power law, $\tau \propto N^{-\beta}$ with $\beta \approx 0.50 \pm
0.03$.
 \item The failure time $\tau_r$ until a brittle sheet disintegrates into
pieces follows a power law too, $\tau \propto N^{-\phi(f)}$, and an exponential
decay $\tau_r \propto \exp(\mbox{const}/ f^2)$ upon increasing strength of the
pulling force, in agreement with Griffith's criterion for failure.
 \item cracks emerge in the vicinity of membrane edges and typically propagate
parallel to the edges, splitting the sheet in two pieces of size ratio of
$\approx 7\%$.
 \item crack propagation speed is observed to increase rapidly with tensile
force
\end{itemize}

We believe that these findings can be seen as generic also for 2D network
brittle sheets of different geometry (hexagonal lattices, or quadratic lattices
with second nearest-neighbor bonding) where similar interplay between elastic
and fracture behavior is expected to take place. It is clear, however, that
more investigations are needed before a full understanding of fracture in such
systems is achieved.

\section{Acknowledgments}

The authors would like to thank V.G. Rostiashvili for fruitful discussions.
A. M. gratefully acknowledges support by the Max-Planck-Institute for Polymer
Research during the time of this investigation. This study has been supported
by the Deutsche Forschungsgemeinschaft (DFG), Grant Nos. SFB625/B4 and FOR597.
H.P. and A.M. acknowledge the use of computing facilities of Madara Computer
Center at Bulg. Acad. Sci.

\end{document}